# The lure of misleading causal statements in functional connectivity research

David Marc Anton Mehler[1, 2] and Konrad Paul Kording[3, 4]

[1] Cardiff University Brain Research imaging Centre (CUBRIC), School of Psychology, and MRC Centre for Neuropsychiatric Genetics and Genomics, Division of Psychological Medicine and Clinical Neurosciences, School of Medicine, Cardiff University, Cardiff CF244HQ, UK

[2] Department of Psychiatry, University of Münster, Germany

[3] Department of Bioengineering and Department of Neuroscience, University of Pennsylvania, USA

[4] Canadian Institute for advanced Research (CIFAR)

**Abstract:**

As neuroscientists we want to understand how causal interactions or mechanisms within the brain give rise to perception, cognition, and behavior. It is typical to estimate interaction effects from measured activity using statistical techniques such as functional connectivity, Granger Causality, or information flow, whose outcomes are often falsely treated as revealing mechanistic insight. Since these statistical techniques fit models to low-dimensional measurements from brains, they ignore the fact that brain activity is high-dimensional. Here we focus on the obvious confound of common inputs: the countless unobserved variables likely have more influence than the few observed ones. Any given observed correlation can be explained by an infinite set of causal models that take into account the unobserved variables. Therefore, correlations within massively undersampled measurements tell us little about mechanisms. We argue that these mis-inferences of causality from correlation are augmented by an implicit redefinition of words that suggest mechanisms, such as connectivity, causality, and flow.

**The state of the estimated connectivity field**

A fundamental goal in neuroscience is to understand brain mechanisms that underlie perception, cognition, and behavior, which, arguably, requires understanding the causal interactions between neurons and neuronal populations. Whenever we talk about causal interactions in this opinion paper, we use the counterfactual definition. A variable causally influences another variable if a perturbation of the first variable would induce a change in the activity of another variable (Pearl, 2009a). This nicely approximates what we mean as neuroscientists: if we say a neuron influences another neuron, we mean that perturbing the first (say electrically) would affect the second and if we say that a region influences another we mean that perturbing the first region (say magnetically) would affect the second. Hence, causality has an intuitively clear definition (counterfactuals or perturbation) and we should demand our statistical approaches to be measured against it (although see also Gomez-Marin, 2017).

Because we cannot directly measure these interactions, statistical techniques are used that aim to infer interactions from simultaneously recorded brain signals. This often boils down to asking the question of how does neural population A mechanistically affect population B? The approaches that we call estimated connectivity (eC) in this paper convert measured signals into a statistical estimate of "connectivity." The results are often (implicitly or explicitly) thought of as a measure or at least approximation of the causal strength of interactions. A rich body of literature has described eC techniques: some techniques utilize granger causality (Bressler & Seth, 2011), other techniques talk about *information flow* (Babiloni et al., 2005; Honey, Kötter, Breakspear, & Sporns, 2007). Another class more directly talks about Dynamic *Causal* Modeling (Daunizeau, David, & Stephan, 2011). Within the imaging community the term *effective connectivity* (EC) is often used when causality is explicitly claimed but within our definition, they fall into our more expansive definition of eC. We will argue that the used observational approaches share the same logical weakness – statistical confounding. Yet, the lure of extracting causality from observational data, is so powerful, that we cannot avoid feeling the pull of it and have effectively referred to correlations in connectivity terms (Stevenson, Rebesco, Miller, & Körding, 2008).

The problem of unobserved confounding is the existence of unobserved variables that affect the observed variables in a way that often makes the estimation of connectivity impossible. It is easy to see why it is impossible to use any statistical techniques to estimate causal interactions in the presence of unobserved confounding. Let us say there is a causal influence from A to B. In this case, it is always possible to construct a common input C which will produce the same effect on B that A would have (e.g. by replicating A and feeding into C). Similarly, if there is no influence from A to B it is always possible to use a confounder to change A and B so that they now look like they do interact, according to any algorithm used. These issues would maybe be a minor problem if we had reasons to believe that confounding is weak, i.e. if there were not many orders of magnitude more confounders than measured variables. However, we will demonstrate that we have no reason to believe so illustrating why it is highly unlikely that an algorithm could identify the network of interactions.

Techniques used for estimating connectivity typically come from fields that focus on forecasting or describing time-series. For example, Granger Causality (Bressler & Seth, 2011) and Coherence analysis (Bastos & Schoffelen, 2016; Sun, Miller, & D'Esposito, 2004) have been developed to describe the relationship between signals. They were then translated to address biological questions such as "How does neural population A interact with neural population B?" (Bressler & Seth, 2011; K. J. Friston et al., 1997; K. J. Friston, Harrison, & Penny, 2003). In many cases, scientists simply analyze the correlations and take high correlations to be a sign of neuronal *communication*. Alternatively, delayed or nonlinear correlations are used to estimate the direction of *interactions*. However, in biological questions we usually seek causal mechanisms and not just good predictions. Hence, researchers increasingly use, in our view without much discussion of threats from confounding, techniques that were used to describe the relationships between signals to make claims about causality.

Indeed, many scientists have been developing techniques for estimating the strength of connections between neurons based on simultaneous spike recordings. The underlying idea is that we want to predict each neuron's spiking probability based on the activity of other observed neurons (Pillow et al., 2008). And indeed, if we record all neurons, they are noisy, and we know that from their transfer function we should be able to estimate the strength and nature of causal interactions (Karbasi, Salavati, & Vetterli, 2018). There has been ample speculation about the meaning of the results of such a study, but it is frequently interpreted in causal terms (Pillow et al., 2008; Stevenson et al., 2008).

Another group of scientists has been developing techniques for estimating the strength of interactions between brain areas using neuroimaging techniques such as Electroencephalography (EEG), Magnetoencephalography (MEG), and functional magnetic resonance imaging (fMRI). Estimated connectivity (eC) in neuroimaging has been intensively studied with data acquired during tasks, as well as rest (resting-state connectivity). Within the eC community one can delineate two different philosophies. Some authors use eC to describe replicable network properties of functional and anatomical neural data without attributing causal significance to interregional correlations (Raichle et al., 2001; Smith et al., 2009; Yeo et al., 2011). Others interpret changes in parameter estimates in more explicitly causal terms such as contributing factors of pathophysiological processes (Hacker, Perlmutter, Criswell, Ances, & Snyder, 2012; Karlsgodt et al., 2008; Pawela et al., 2010; Wu et al., 2009), biological mechanisms in learning and cognition in healthy (Danielle S. Bassett, Yang, Wymbs, & Grafton, 2015; Nauhaus, Busse, Carandini, & Ringach, 2009; Siegel, Donner, & Engel, 2012; Stevenson et al., 2008), or in terms neuroplasticity in the functional re-organization of the brain (Hallam et al., 2018; Jin, Lin, Auh, & Hallett, 2011; Lueken et al., 2013). Indeed, recent bibliometric research suggests that FC is a substantial area in neuroimaging (Dworkin, Shinohara, & Bassett, 2018), and it is thus is important to obtain clarity about its interpretation.

**It is not just semantics?**

The eC measures have in common that they are functions of (sometimes generalized) correlations. However, the language used generally does not reflect that. Almost every term in popular use suggests causality. For example, we use terms like Granger *Causality*, functional *connectivity,* information *flow*, *effective connectivity*, dynamic *causal* models, etc. We do this despite the fact that other terms such as *improvement in predictive power*, *correlations*, *conditional correlations*, and *model comparison* can denote more precisely what we actually do. To deal with the disagreement between used correlational techniques and desired mechanistic or causal statements, we are effectively redefining the English language. Connectivity implies a connection between two places. Causality implies cause and effect. Flow implies that something moves from one place to another. Effective implies that something has an effect. This set of re-definitions gives rise to the problem that eC approaches are often misunderstood.

Scientists write about connections within the brain minimizing the wiring length along which signals need to travel (E T Bullmore & Sporns, 2009; Edward T. Bullmore & Bassett, 2011), but while the brain may want to minimize the length of its wires, there is no implication that it pays any price for correlations. They write about stimulation to control the network (Taylor et al., 2015), which requires interactions to be causal. They write about interference to cure diseases (Khambhati, Davis, Lucas, Litt, & Bassett, 2016), which again requires causality. Or they write about regions that "cause more exchange of causal information" (Bajaj, Butler, Drake, & Dhamala, 2015). These examples show how correlations are assumed to indicate causality. Specifically, eC is often used as if it did reveal an approximate understanding of causality, and much of it is due to misleading use of words that imply causality in lay English and merely refer to algorithm properties in statistics.

For example, we ourselves wrote in 2008 " […] [estimated connectivity] methods have become staples of neural data analysis, and have revealed a great deal about the interactions between cortical and subcortical structures." (Stevenson et al., 2008). We could simply have said that models that use other activities as independent variables are good predictors. With LFPs it was argued that "[…] the relative weight of feedforward and lateral inputs in visual cortex is not fixed, but rather depends on stimulus contrast." (Nauhaus et al., 2009). For EEG and MEG power coherence analyses, it was advocated that "[…] amplitude correlation is an informative index of the large-scale cortical interactions that mediate cognition." (Siegel et al., 2012). For fMRI data, other authors reported "[…] changes in the architecture of functional connectivity patterns that promote learning from initial training through mastery of a simple motor skill." (Danielle S. Bassett et al., 2015). However, all these approaches only reach level one in Pearl's hierarchy of causation (Dablander, 2019; Pearl & Mackenzie, 2018). While the impossibility of making causal statements in such situations is well known (Dawid, 2008; Holland, 2015), the field regularly makes causal statements based on statistical analyses that cannot support such statements. Altogether, we conclude that this debate is not merely about semantics and second that "whereof one cannot speak, thereof one must be silent" (Wittgenstein, 1922).

**We do not actually learn about causality from estimated connectivity**

Here we ask if the eC techniques used should be expected to estimate causal interactions. We ask if the techniques measure causality, connectivity, or flow, in the standard meanings of these words. We will conclude that, due to massive confounding, they merely describe the statistics of the joint neural data without convincing causal insights.

Neuroimaging datasets from fMRI and MEG are affected by obvious confounders including head motion and physiological processes such as the heartbeat and breathing (Driver, Whittaker, Bright, Muthukumaraswamy, & Murphy, 2016; Messaritaki et al., 2017; Murphy, Birn, & Bandettini, 2013) that add structured noise to the signal. In response, the field has developed some effective data correction techniques (Ciric et al., 2018, 2017; Murphy et al., 2013; Parkes, Fulcher, Yu¨cel, & Fornitod, 2017; Power et al., 2014). One dominant view is that correcting for such obvious confounders, for instance by using partial correlation analyses, improves causal inference irrespective of unobserved confounders(Reid et al., 2019). However, for each obvious external controllable confounder there are countless internal unobservable ones (DiDomenico & Eaton, 1988). Hence, while we agree that correcting for observed confounders improves the prediction of models, we argue that such view remains speculative when it comes to inferring causality for theoretical reasons that we lay out below.

A growing field in the domain of statistics and econometrics is concerned with causal discovery. They ask how observational data can be used to answer causal questions. There are multiple philosophies. Some base their studies on fitting modern techniques such as directed acyclic Bayesian graphs that represent a set of variables and their conditional dependencies (Pearl, 2009a). Others focus on matching approaches that can account retrospectively for randomization (Rubin, 2010). Importantly though, both have in common that they fail if there are important variables that are unmeasured and affect the variables of interest. The effect induced by other variables on the apparent statistical relation between variables of interest is called confounding. Within observational causal inference there is an understanding that there can be no solution if each pair of measured variables shares an unobserved confounder (Pearl, 2009b).

A lot of the discussion in the field of eC is about statistics. And indeed, statistically estimating causal interactions is a hard problem (Spirtes & Zhang, 2016). Over time, advocates of functional connectivity research have introduced several statistical innovations, including sophisticated generative models for spiking (Havlicek, Ivanov, Roebroeck, & Uludağ, 2017; Paninski, Brown, Iyengar, & Kass, 2008), priors about neuronal interactions (Calabrese, Schumacher, Schneider, Paninski, & Woolley, 2011; Stevenson et al., 2009), and priors about the network of interactions (D. S. Bassett, Meyer-Lindenberg, Achard, Duke, & Bullmore, 2006). This shows that the statistics of network inference is complex, reflected by a progressively sophisticated field. However, we argue here that it is impossible to overcome the problem of massive confounding by using statistics.

We want to dwell on the idea of confounders to eC. Let us say we are interested in the interactions between two neurons. Further, let us assume that there is a third neuron (confounder) that activates both neurons. In that case, if we see a correlation between the neurons we cannot know if it is due to the neurons interacting directly or through an induced interaction by the unobserved neuron. More generally, any joint distribution between two neurons could be induced by a single third neuron. Unobserved neurons or information can act as confounders for the inference of eC, threatening the validity of the results. This confounding problem is central to the statistical field of *causal inference* and it is generally acknowledged that, in typical situations, unobserved confounders render inferences about causality impossible.

We want to use the Simpson's paradox to highlight the threat of confounding (Simpson, 1951), in which unobserved interaction can either cancel out main effects, or artificially induce spurious main effects. Let us say that there are two brain states (e.g. related to two levels of attention), and that one state is associated with high activity of neuron A and low activity of neuron B while another is associated with low activity of neuron A and high activity of neuron B. But let us say that there is a positive instantaneous causal influence from one on the other. If we do not know the brain state (confounded) we may conclude that neuron A has a negative influence on the activity of neuron B. But if we do know the brain state we correctly see that A increases the activity of neuron B (Figure 1). This paradox shows how confounders with relevant structure can arbitrarily influence the resulting eC.

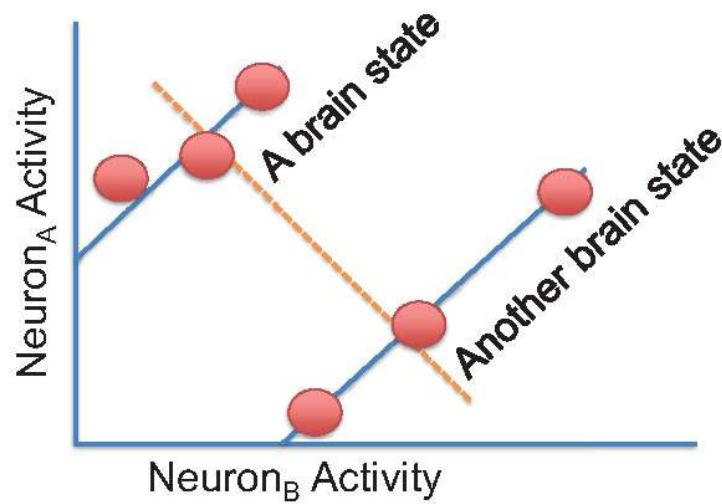

Figure 1    The red dots represent neural activity recordings of neuron A and B. The yellow line indicates a negative correlation between activity of neuron A and B. However, this correlation is confounded by two different brain states. The blue lines indicate that within each brain state there is a positive correlation between the activity of neuron A and B. To unravel this positive correlation, knowledge about the different brain states is required. The Simpson's paradox shows that the sign of a correlation is not indicative of causality.

Indeed, in certain neuroimaging datasets it is possible to directly observe the Simpson's paradox. For instance, the choice of analysis parameters, such as seed regions, and the network they belong to, may mediate whether different FC methods yield similar, or orthogonal results (Roberts, Hach, Tippett, & Addis, 2016). Moreover, functional brain parcellations reconfigure based on the brain state (Salehi et al., 2019).

**Massive confounding destroys the causal interpretation of estimated connectivity**

Our argument about the impossibility to obtain causality from eC does not come from data – there is too little ground truth about complex brains to make this possible (Jonas & Kording, 2017). It rests on a simple consideration of the factors that are known to make causal inference theoretically impossible. It rests on the justified assumption that we record only few of the neurons when recording spikes, or a few projections of neural activities in imaging. We have no convincing reasons to assume that the observed dimensions should be more important than the unobserved. Our argument thus boils down to a simple logical statement: if the bulk of causal interactions happen from and within unobserved dimensions, then the correlations between observed variables are simply epiphenomena. Correlation is not causation, regardless the mathematical sophistication we use when calculating it. Causal inference algorithms that work with observational data are generally built on the assumption of causal sufficiency, which boils down to there being no unobserved confounders (although see Ranganath and Perotte, 2018; Wang and Blei, 2018). Without these assumptions we can at best produce families of potential models and if any pair of recorded variables is confounded then this family will contain all models (Spirtes, Glymour, & Scheines, 2001). Recording only a few variables in a densely interacting causal system generally renders causal inference impossible (Jonas Peters, Dominik Janzing, & Bernhard Scholkopf, 2017; Pearl, 2009a).

    We can be more precise about the exact reasons for the failure to meaningfully obtain causality (Pearl, 2009a). Within causality it is known that, under certain conditions, causality can be established. Causality from A onto B can be primarily established if one of two conditions are

given. (1) The back-door criterion: We observe a set of variables S that blocks all causal influences onto A and no node in S is descendent of A. (2) The front-door criterion: We observe a set of variables S that block all direct causal paths from A to B, there are no unblocked backdoor paths from S to A, and A blocks all back-door paths from S to B. In other words, we would either need to measure all variables that affect a variable of interest which is generally not possible) or, probably harder, we would need to record all variables that could mediate the influence of A on B. Both sets S will, for many typically questions in neuroscience, include millions of neurons.

When analyzing spike data, there are far more unobserved variables than observed variables. When we record a few hundred neurons (Stevenson & Kording, 2011), the number of recorded neurons is a vanishingly small subset of all neurons. We have little reason to assume that the recorded neurons are much more important than the un-recorded neurons. As each neuron receives inputs from so many other un-recorded neurons, we should expect that the parts of neural activity driven by unobserved neurons are arbitrarily larger than the parts coming from observed neurons. In other words, the confounding signal should be many orders of magnitude more important than those coming from observed data. As such, we should not expect that causal inference is possible.

When analyzing imaging data such as fMRI, or LFP, EEG, or MEG, there are also far more unobserved variables than observed variables. Within each signal source, we can, in some abstraction, observe the sum of neural activity. But the same measured activity can be realized by any combination of individual activities rendering a solution of the inverse problem (signals → neuronal spike trains) infeasible. The activity of neurons which are orthogonal to our signal, can span arbitrary dimensions, related to movement, memory, thought or neuronal communication. Importantly, dense physiological recordings in small areas suggest that countless variables are represented (e.g. movement related signals in V1; Musall et al., 2018; Stringer et al., 2018). The signals that we can measure are arbitrarily low-dimensional relative to the full dimensionality of communication in the brain. As such we are still in the situation where we have a number of confounders that is many orders of magnitude larger than the number of measured variables. This again puts us into the domain where causal inference should be impossible.

We might feel that causality may happen at a given scale, rendering the argument about unobserved dimensions invalid and allowing a multi-scale definition of interactions. The argument given is often the analogy to statistical physics: while understanding the interaction between gas molecules is hopeless, a large set of atoms can be well understood in terms of temperature and pressure. However, this analogy quickly breaks down. Every neuron is special in the sense that they do not interact with random neurons, but with a largely fixed set, or worse, a set that changes through neuroplasticity. The justification of averaging over molecules is often perfectly fine in statistical physics. However, whether this this logic works in neuroscience remains an open question.

One may argue that there are a large number of studies that use mean field approaches (Deco, Jirsa, Robinson, Breakspear, & Friston, 2008; K. Friston, 2008; Moran, Pinotsis, & Friston, 2013; Pinotsis, Robinson, Graben, & Friston, 2014; Wong & Wang, 2006). It is argued that "because each neuron receives input from many others, it is sensitive only to their average activity ("mean field")" (Gerstner, Sprekeler, & Deco, 2012). However, these studies set up simulations that unfold in low-dimensional spaces and then show that they can be recovered in low-dimensional spaces. To our knowledge no study has shown that networks that compute like brains can be approximated by the underlying mean-field approximation. Barring such a study we should assume that mean-field is not a meaningful approximation of brain computation. We therefore conclude that the separation of scales that sounds so plausible outside of the brain makes little sense within.

**Why (somewhat) higher resolution is not the solution**

As the impossibility of causal inference from subsampled data feels counter-intuitive we want to spell out the problem a bit more. Let us assume we are interested in the connections between two signals, e.g. voxels B and C. But the brain's real dynamics is characterized by the activity of all neurons ($x_t$). Let us assume, for simplicities sake, that dynamics are linear:

**$x_t = Ax_{t-1}$ + noise**     (1)

where matrix **A** describes the true dynamic of the system. The activities in our original voxels of interest, B and C (conveniently concatenated into a vector **y** which is actually observable) can be obtained using a projection matrix **M** that projects the activity of all neurons into a two-dimensional space:

**$y_t = Mx_t$.**     (2)

A neuroscientist may then want to use the time lagged correlation **R**=< $y_t$ $y_{t-1}^t$> in the measured signal **y** to gain insights into the properties of **A**. What will we measure then? We can now insert the definition of **x** and **y** and obtain

**R=MAVM$^t$**     (3)

where V is the covariance matrix of x. The question now is what we can learn about **A** from **R**. We can obtain some intuition in special cases. If **V** was the identity matrix and hence all neurons would fire entirely independent of one another then **R** would reveal the average influence of brain area B on C, only that then the connections would have all to be zero. However, in reality, the autocorrelation function in every brain area (the local **V**) is known to have a broad spectrum of singular values. Moreover, **V** between brain areas will be nonzero. Here, confounding becomes obvious: elements of **A** that neither relate to signals B or C can in arbitrary ways affect the correlation **V** between the signals (see Figure 2). Importantly, if N is the number of neurons and K the number of measured signals (two in the shown example) there is an $N^2 - K^2$ manifold of **A** matrices that produce an identical correlation matrix. As the number of neurons is typically much larger than the number of measured signals, the measurements do not really lower the dimensionality of the space of potential models ($N^2 - K^2 \approx N^2$). We note that while in this example we have computed a time-delayed correlation matrix as a metric of eC, also different methods (e.g. in Granger causality) essentially suffer from the same problem.

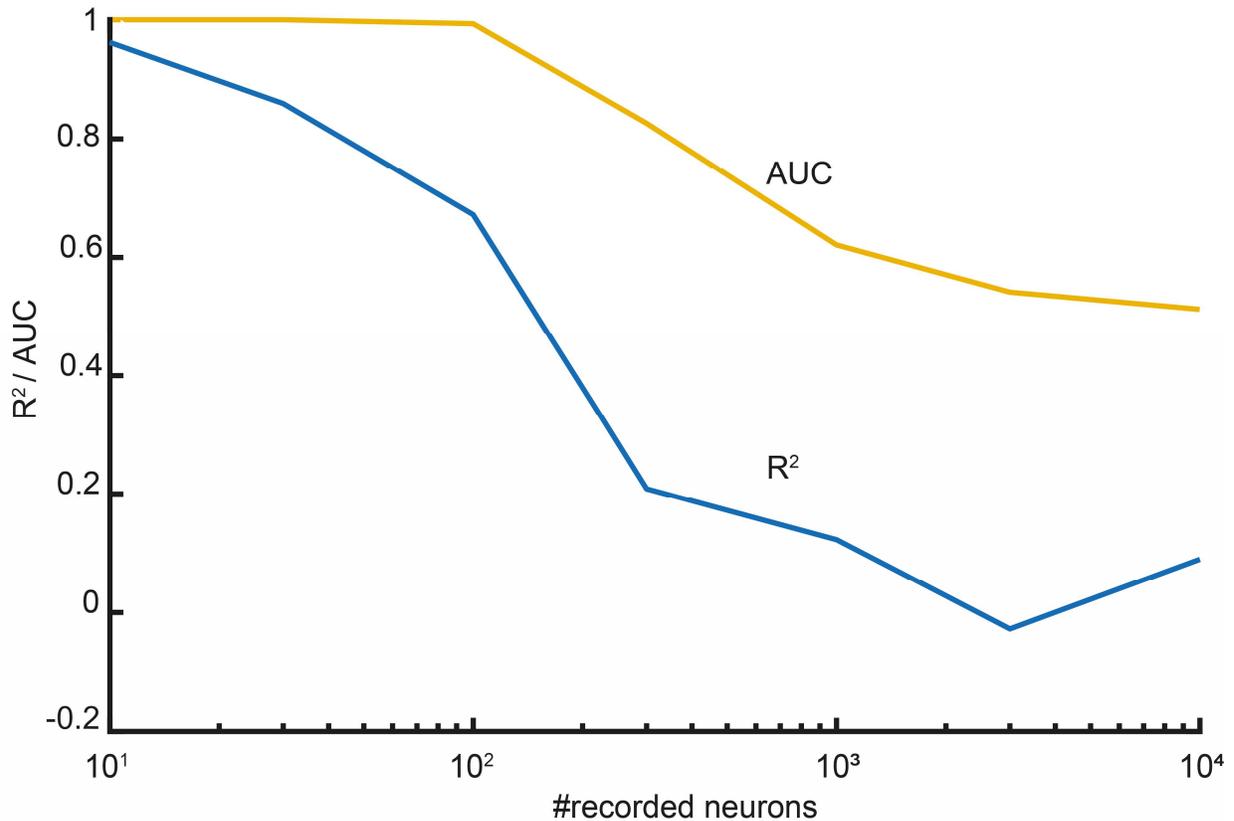

Figure 2    Correlation is not causation for big networks. To quantify the similarity of correlation and causation as a function of the network size we implemented a network with varying number of neurons (x-axis) and binary weights. We plot the $R^2$ in yellow and the area under the curve (AUC) in blue, both measures of prediction ability, when using the correlations as an estimator for the causal strength. Network reconstructions rapidly decreases in quality as the network gets larger.

As the neural dimensionality (or speed of processing) in each measured signal increases, the temporal resolution decreases, the noise (or non-communication related signal) increases and hence the idea of extracting causality from observation becomes hopeless. Importantly, this is not a problem that can simply be solved by recording data from more subjects. This is a case where a problem can fundamentally not be solved.

For all methods that we have at our disposal to run functional connectivity analyses, the signals are few relative to the many unobserved variables. In these settings, interactions between measured signals should describe much less variance than the interactions between unobserved variables. Just like in Simpson's paradox, the interactions between such unobserved variables can arbitrarily affect the estimates of interactions between measured signals. In neuroscience we are essentially always in the massively confounded situation and then correlations as not informative about causality.

**eC fails basic tests of empirical causality**

So far, we have reviewed the logical evidence why eC from a small number of channels should, due to massive confounding, not be a suitable tool to reveal causal structure of brains that contain billions of neurons. However, in principle there may be aspects of brain activity that could rescue the idea. For example, if the brain's activity was very low dimensional (Cunningham & Yu, 2014; Yu et al., 2009), then recording from a small number of voxels or neurons may be equivalent to recording all of them. However, in this case all the neurons that jointly define a dimension will confound the causal inference. Similarly, if we believe that the composition of the activity within a voxel does not matter but just the sum of the activities, and thus subscribe to a strong mean-field view (Cooper & Scofield, 1988; Gerstner et al., 2012), the statistical problems may be resolved. However, within each voxel, we find neurons of countless tuning properties (Hubel & Wiesel, 1962). Alternatively, there may be something informative in the structure of neuronal signals that somehow makes this analysis possible. We may hope to gain some insight into the truth by analyzing the algorithm performance in situations where we know what to expect.

If we should expect eC to meaningfully work then we should, above all, expect that it would work in simulated situations that lack many of the complexities characterizing the real world. And indeed, a recent study has systematically analyzed the quality of eC is based on the number of neurons that were not simulated. Removing as many as 20% of the neurons from the recordings let the reconstruction quality drop from 100% to 70% (with a chance level of 50%). Hence, eC measures can tolerate some proportion of missing neurons but only a small proportion (compare to nearly 100% for typical spike analysis; (Karbasi et al., 2018)). Further, eC in spiking data is highly susceptible to effects of temporal smoothing (Stevenson & Körding, 2010). Hence, spatial and temporal smoothing in order that are routinely applied can distort eC substantially, casting doubt on the use of such techniques for inferring causality at the neural level.

**Effective connectivity – a special case?**

Lastly, parts of the eC community, those using the term *effective connectivity* (EC) introduced a rather interesting philosophical alternative. In this view EC produces, or *causes* observed eC in a mathematical sense (Karl J. Friston, 2011). As such they are causal given unverified, and arguably unlikely, assumptions. Interestingly, DCM derives from ideas of perturbations, it conceptualizes stimuli as perturbations (K. J. Friston et al., 2003). However, this strategy equally cannot guard against the potential effects of confounding. For instance, "the "missing region"" – i.e. underdetermination problem – presents has been referred to as "a potential deep methodological confound" by DCM proponents (Daunizeau et al., 2011). Indeed, *neurostatisticians* do not explicitly claim that they are studying real connectivity, but rather perform hypothesis-based model comparison and tests on estimated parameters, allowing to them make statements about the preferable model in a given model class. The statements afforded by this approach are thus not about (biological) causality in the brain but about preference for a model from that model class ("All models are wrong, but some are useful") (Box & Draper, 1987). Yet, once Dynamic Causal Modeling (DCM) is applied to brain signals, it is claimed that these models "estimate neurobiologically interpretable quantities such as the effective strength of synaptic connections among neuronal populations and their context-dependent modulation" (Stephan et al., 2010). Thus, DCM applications that implicitly aim to unravel neural mechanisms misinterpret causality in the model as biological causality in the same way as other eC techniques. We further note that expanding the model class can arbitrarily change the resulting causal conclusions. This approach does thus not afford statements about causal interactions within the brain; it rather risks falling for internally consistent statements about network dynamics that may be causal in a mathematical, but not in a biological sense (Etkin, 2018).

**The potential for correct "estimated connectivity" in the future**

Algorithms claiming eC have been used exhaustively when analyzing brain data with the hope of getting at causality. Using a more accurate terminology could help in the interpretation and, clarify that our results remain foremost descriptive: *Interregional*, or *interneural signal correlations* captures what most techniques measure.

For instance, it is possible to perturb the brain in many ways and thus to go beyond purely correlational experiments. To confidently get at causality requires perturbations to evaluate how a given input to the system modulates activities. Invasive brain stimulation techniques (e.g. optogenetics, intracortical electrical stimulation, and deep brain stimulation) and non-invasive brain stimulation techniques (e.g. transcranial magnetic stimulation and transcranial electrical stimulation) allow brain perturbations. These techniques are currently, despite their limitations (Bestmann, de Berker, Bonaiuto, Berker, & Bonaiuto, 2015; Häusser, 2014; Sack & Linden, 2003; Siebner, Hartwigsen, Kassuba, & Rothwell, 2009), as close as we can get to causality in neuroscience (Chen & Rothwell, 2012; Muldoon et al., 2016). However, we also note that inference about the inverse is more complex: lack of behavioral response after perturbing a certain area does not a imply that it is not causally but may be merely due to compensatory recruitment (Krakauer, Ghazanfar, Gomez-Marin, MacIver, & Poeppel, 2017; O'Shea, Johansen-Berg, Trief, Göbel, & Rushworth, 2007; Sack & Linden, 2003).

Simulated perturbation experiments allow us to check if our assumptions are correct and what happens if they aren't. Computational studies are an extension of thought experiments and have a long tradition in neuroscience (Gerstner et al., 2012; Lapicque, 1907). For example, we can evaluate the susceptibility to pitfalls of various common eC (Bastos & Schoffelen, 2016) and test how low signal-to-noise ratios affect false positive rates. Another pitfall of non-invasive electrophysiological studies (EEG/MEG) is volume conductance, which can lead to spurious eC, and thus require sanity checks (Haufe, Nikulin, & Nolte, 2012). For the obvious confounders the field is starting to use simulations to test basic assumptions of their techniques (Bastos & Schoffelen, 2016; Bright & Murphy, 2015; Haufe, Nikulin, Müller, & Nolte, 2013; Ramsey et al., 2010; Rangaprakash, Wu, Marinazzo, Hu, & Deshpande, 2018; Seth, Chorley, & Barnett, 2013; Stokes & Purdon, 2018; Thompson, Richter, Plavén-Sigray, & Fransson, 2018). We thus propose that *Interregional*, or *interneural signal correlations* should be routinely checked for obvious signs of the Simpson's paradox as has been suggested for other fields (Kievit, Frankenhuis, Waldorp, & Borsboom, 2013; Rousselet & Pernet, 2012; Tu, Gunnell, & Gilthorpe, 2008). Simulation will help us to validate approaches, test the sensitivity of sanity checks and the effectiveness of potential remedies.

Other (mostly non-invasive) perturbation techniques such as neurofeedback training and brain computer interfaces (BCIs) may be seen as an approximation to direct (self-) control in humans (Arns et al., 2017; Chaudhary, Birbaumer, & Ramos-Murguialday, 2016; Sitaram et al., 2017; Watanabe, Sasaki, Shibata, & Kawato, 2017). These allow to entrain correlations between brain regions and test for desired behavioral changes (Ramot et al., 2017; Yamashita, Hayasaka, Kawato, & Imamizu, 2017). Similarly, BCIs allow coupling neural firing rates to behavioral outcomes and thus they may provide new tools for trying to understand causality within the brain (Golub et al., 2018; Sadtler et al., 2014). However, such complex interventions entail various confounders and require experiments with (often many) carefully designed control conditions (Sorger, Scharnowski, Linden, Hampson, & Young, 2019). Invasive closed-loop optogenetic stimulation in animals, where stimulation is triggered by learned neural activity patterns may circumvent many of these confounders and allow studying causality (Athalye, Santos, Carmena, & Costa, 2018). Of interest, combining invasive optogenetics with non-invasive neuroimaging yields new possibilities to test for causal *effective connectivity* (Bernal-Casas, Lee, Weitz, & Lee, 2017).

For the analysis of some spiking systems there may be ways of sidestepping the confounding problem. If we have fast recordings and sparse connectivity patterns so that there is no time for the signal to travel sufficiently quickly from one neuron to another through any path but the most direct one, then the system effectively gets to be unconfounded. While this is not the typical setting, there may thus be ways of sidestepping it for certain subproblems (Bartho, 2004; English et al., 2017; H. A. Swadlow & Giisev, 2001; Harvey A. Swadlow & Gusev, 2002; Usrey, Alonso, & Reid, 2000). The intuition here is that if we record sufficiently fast, we see the direct and immediate effect of one neuron on another such that we can effectively use the timescale to *de-confound* estimates.

It may also be possible that there is something about brain signals that, given the right analysis methods, makes it possible to approximately estimate causal interactions, for instance between brain regions. Maybe there are signals that are low dimensional and localized, maybe there are sparse localized events that result in signals that approximate external perturbations. For example, quasi-experimental techniques may allow meaningful causal estimates about brain connectivity (Lansdell & Kording, 2018; Lepperød, Tristan, Torkel, Marianne, & Kording, 2018). However, we cannot meaningfully use such techniques to learn about causality 1) without having established that the strong extra assumptions we would need to make about the brain are justified and 2) before having used simulations to check that under those assumptions the methods would actually work. The extensive confounding from only observing a small subset of neurons, however, seems hard to overcome without massive scale recording technologies or small brains. Many algorithms used for the inference of functional connectivity are quite meaningful when applied to systems that are exhaustively recorded. As such, it is an interesting question how well they work on small animals or engineered systems. For example, in the worm c. elegans, aplysia, or microprocessors (Jonas & Kording, 2017) recording all "neurons" at high temporal precision should be possible and in the larval zebrafish that may be possible soon (Ahrens, Orger, Robson, Li, & Keller, 2013). Preliminary work that compared connectivity matrices of functional and structural whole-brain data found little overlap between these (Saunders, 2019). In systems where the complete circuitry is described, perturbation (e.g. pharmacological) studies can be used to test whether algorithms can reconstruct expected patterns (Gerhard et al., 2013). Such systems are much closer to the implied assumptions of the various causal inference techniques, mainly because they make the confounders observable. Although there would obviously still be confounding, e.g. from limited temporal resolution, making the causal inference problem quite hard, promising approaches from econometrics, e.g. regression discontinuity designs and other pseudo experiments (Angrist & Pischke, 2008) may be helpful (Marinescu, Lawlor, & Kording, 2018). Nevertheless, even if all neurons were recorded in such systems, the causal inference problem at scale would remain outstandingly difficult (Das & Fiete, 2019).

We have argued here, that eC approaches cannot meaningfully get at causality due to massive confounding, but it is important to point out that other branches of neuroscience and biology more generally have the same problem (Jonas & Kording, 2017). For instance, tuning curves, which describe how neurons are affected by a stimulus dimension such as color, do not reveal how they are computed. Lesions usually induce compensation making it hard to interpret their causality. Pharmacological interventions or stimulation studies typically perturb many neurons making it hard to assign the undeniable causal link to any specific neuronal path. However, the logical problems in other areas of neuroscience do not render a lack of logical precision in the eC field (more) acceptable. It is time for the field to take the reality of causal inference seriously (Angrist & Pischke, 2008; Jonas Peters et al., 2017; Pearl, 2009a).

**Understanding the joint statistics of neurons is still interesting**

We have reviewed why eC approaches cannot meaningfully get at causality due to massive confounding from unobserved variables. However, this circumstance does not imply that trying to understand the joint dynamics of many neurons or brain areas is not interesting. For example, looking at the brain in lower dimensional projections may allow us to see its invariances (Bruno, Frost, & Humphries, 2017; Gallego et al., 2017; Gordon et al., 2017). *Interregional correlations* may have no causal meaning, but they may allow us to derive biomarkers (Beijers, Wardenaar, van Loo, & Schoevers, 2019; Desowska & Turner, 2019; Marceau, Meuldijk, Townsend, Solowij, & Grenyer, 2018). To which degree these reflect neuronal, however, remains subject to future well controlled studies. Informative markers may even be derived from brain-body correlations (Rebollo, Devauchelle, Béranger, & Tallon-Baudry, 2018; Valenza, Toschi, & Barbieri, 2016). In fact, there are many biologically meaningful questions (Gomez-Marin, 2017; Krakauer et al., 2017) – and insights about brains can come from answering any of them. It is just important to be clear about the statements permitted by any one approach – statements about joint statistics are not meaningful statements about causality.

Along with these considerations comes an important set of insights into the sociology of neuroscience. The current publication system almost forces authors to make causal statements using filler verbs (e.g. to drive, alter, promote) as a form of storytelling (Gomez-Marin, 2017); without such a statement they are often accused of just collecting meaningless facts. In the light of our discussion this is a major mistake, which incites the field to misrepresent its findings. Understanding the structure of brain data is interesting in its own right. Scientists who actually measure causality using carefully designed perturbations should be lauded for the hard work. At the same time, scientists who describe joint statistics should be rewarded for careful characterizations. We do learn about the brain by analyzing joint distributions. We simply should not claim causality.

**Acknowledgements**

We would like to thank our colleagues for insightful comments and discussions that have substantially contributed to the content of this manuscript: Alex Gomez Martin, Anil Seth, Brian Collopy, Daniele Marinazzo, David Linden, Dimitris Pinotsis, Ian Stevenson, Jean Daunizeau, Johannes Algermissen, Jonathan Pillow, Juan Alvaro Galego, Karl Friston, Klaas Enno Stephan, Lionel Barnett, Manjari Narayan, Michael Haufe, Robert Kass, Roscoe Brady, Russ Poldrack and Stefan Rotter (in alphabetical order).

**Data availability statement**

Data and code available on: https://osf.io/9cs8p/

**Conflict of interest**

DMAM receives payments to consult with a neurofeedback start-up company. KPK has no conflict of interest to declare.